\documentclass[aps,prb,twocolumn,groupedaddress]{revtex4}

\usepackage{graphicx}
\newcommand{\be}{\begin{equation}}
\newcommand{\ee}{\end{equation}}

\begin{document}
\title{Peas in a pod: quasi-one-dimensional C$_{60}$ molecules 
in a nanotube}
\author{M. Mercedes Calbi, Silvina M. Gatica and Milton W. Cole}
\affiliation{Physics Department, Pennsylvania State University, University Park,\\
Pennsylvania 16802}
\date{\today }

\begin{abstract}

\vspace{.5cm}

\begin{center}
\begin{tabular}{r}
{\em ``How luscious lies the pea within the pod...''} \\
{\em Emily Dickinson}
\end{tabular}
\end{center}

\vspace{.5cm}

We evaluate the equation of state of the quasi-one-dimensional 
(1D) phase of C$_{60}$ molecules in small carbon nanotubes, 
nicknamed ``peas in a pod''. The chemical potential and 1D 
pressure are evaluated as functions of the temperature and 
density, initially with the approximation of nearest neighbor 
interactions and classical statistical mechanics. Quantum 
corrections and long-range interaction corrections are discussed, 
as are the effects of interactions with neighboring peapods. 
Transition phenomena involving the 3D coupling are evaluated.

\end{abstract}

\pacs{}
\maketitle

\section{Introduction}

Carbon nanotubes have been found to exhibit a wide range of remarkable
properties, concomitants  of   their  nanometer   cross-section  and
macroscopic  length.  Extensive  research  has recently  explored  the
behavior  of gases  adsorbed within,  outside of  and  in-between such
nanotubes \cite{revaldo,boni,boro,rmp,cross,hecond,milenrot}. Among the  most intriguing systems is  that created
when C$_{60}$ buckyballs  lie within a tube of small  radius $R$, since this
is surely a quasi-one-dimensional (1D) system \cite{disc}. This system, nicknamed
``peas in a pod'' has begun  to be studied using various theoretical and
experimental methods \cite{rev,luzzi,template,oth1,univ,1dlatt,size,
thesis,press,vdw}. In this paper, we explore the equation of
state of  this system. Specifically,  we evaluate the 1D  pressure and
the chemical  potential, as functions of temperature  and density. The
calculation is relatively simple in  the small tube case considered in
this paper (e.g. (10,10) tubes,  with $R \approx 7$ $\AA$) because 
the C$_{60}$ molecules
are confined to the vicinity  of the axis and the motion perpendicular
to the axis can  be factored out of the problem, as  we shall see. The
more general case involving tubes  of varying size, with the attendant
appearance  of 3D  effects, requires  a  more detailed  study than  is
described  here.  Some  numerical  studies and  molecular  simulations
exploring this problem have been performed by Hodak and Girifalco 
\cite{1dlatt,size,thesis}.

The next section discusses the model and presents the equation of 
state resulting from the simplest (classical, quasi-1D) procedure 
we have used. Section III assesses quantum and long-range corrections 
to this simple model and estimates the regime of applicability of 
those results. Section IV describes the effects of interactions with 
buckyballs in neighboring tubes, including phase transitions.

\section{Model calculations}

Girifalco and coworkers have shown \cite{univ,size} how C$_{60}$ 
molecules 
within nanotubes of small radius ($R \approx 7$ $\AA$) experience 
potentials confining them to the vicinity of the tube axis. 
This paper employs the potentials developed by that group to determine 
the equation of state of the quasi-1D buckyball fluid. We assume that 
the total potential energy $V_{total}$ of the molecules satisfies

\be
V_{total}(\rho_i,z_i) = \sum_i U(\rho_i) + \sum_{i<j} 
V(|z_i- z_j|)		
\ee

\noindent Here, ${\bf r}_i = (\mbox{\boldmath $\rho$}_i, z_i)$  is the  position 
of  the i-th  molecule's center,
written in  terms of its coordinate $z_i$ parallel to the  tube axis 
and its transverse displacement from the axis, $\mbox{\boldmath $\rho$}_i$. The 
potential $U(\rho)$ is the molecule-tube interaction that  determines the  molecules' 
degree  of localization
near  the axis, which  depends also  on the  temperature $T$.  The other
interaction  in  Eq. (1), $V(z)$, is  that  between  a  pair  of  C$_{60}$
molecules separated a distance $z$ along the tube axis.   
Girifalco et al have shown that both interactions, $U(\rho)$ and $V(z)$, 
can be expressed in a universal way by the following reduced 
function \cite{univ}:

\begin{eqnarray}
\lefteqn{\frac{\Phi(r)}{|\Phi(r_0)|} = } \nonumber \\ 
&& = \frac{5}{3}\left(\frac{3.41}{3.13 
\, {\tilde r}+0.28}\right)^4-\frac{2}{3}\left(\frac{3.41}{3.13 \, 
{\tilde r}+0.28}\right)^{10}
\end{eqnarray}

\noindent where $|\Phi(r_0)|$ is the well depth at the equilibrium 
spacing $r_0$, and ${\tilde r} \equiv (r-r_1)/(r_0-r_1)$. The distance parameter $r_1$ for each system (C$_{60}$-C$_{60}$, C$_{60}$-tube) is 
chosen so that the interactions fit the universal function (2). Table I 
indicates the values of the used parameters.
The function in Eq. (2) is obtained using a continuum model for the carbon
surfaces. The carbon-carbon interatomic potential is integrated over the 
interacting surfaces assuming a mean surface density of carbon atoms and different Lennard-Jones parameters for atoms in C$_{60}$-C$_{60}$ and 
C$_{60}$-graphene systems. Therefore, we ignore any  dependence of the 
interactions on  the atomic structure of  either  the   tubes  or  
the  C$_{60}$  molecules.
Similarly,  the neglect of any interactions  that depend on
the C$_{60}$  orientation means that  the molecular rotational  
problem is not affected by the environment. This means that the 
rotational degree of freedom may be omitted completely from the 
present analysis \cite{milenrot}.

Concerning the C$_{60}$-C$_{60}$ interaction, there  has been  some
discussion in  the literature of the  value of the  asymptotic van der
Waals coefficient \cite{vdw,langreth} ($C_6$, the coefficient of 
$r^{-6}$),  with values  varying over the 
range 16.6 $N_{at}^2$ eV \AA$^{6}$ to 29.05 $N_{at}^2$ eV \AA$^{6}$.
The value used in Girifalco's calculations \cite{univ} 
(Eq. (2)) is 20.0 $N_{at}^2$ eV \AA$^{6}$.

In reality, the intermolecular interaction $V$ ought to depend on the full 3D separation. By neglecting  this interaction's dependence  on 
$\mbox{\boldmath $\rho$}_i$ and  
$\mbox{\boldmath $\rho$}_j$, Eq. (1)
implicitly  assumes that  the molecules' rms  transverse displacement
$d=\sqrt{\langle \rho_i^2 \rangle}$ is small compared to the mean 
separation $a=L/N$ 
along the axis.  We  discuss  the  regime  of  validity  of  that  
approximation
below. Also implicit in Eq. (1) is the assumption that both the nanotube
and  the  C$_{60}$  molecules  are  rigid,  since  otherwise  their  
shape fluctuations  would  affect  the  nanotubes' potential  energy.  
While
conventional in this field,  this assumption deserves future attention
in the case  of such strongly interacting molecules  as C$_{60}$. 

With Eq. (1), the properties of  the system may be evaluated in terms of
independent   transverse  and   longitudinal   contributions  to   the
Hamiltonian.  For  example,  the  partition  function $Z$  of  the  
C$_{60}$
molecules  factorizes into transverse  and longitudinal  factors, 
$Z=Z_t Z_z$, so that the Helmholtz free energy $F$ is an additive 
function of these separate parts of  the problem: $F= F_t+ F_z$. We 
may therefore write the chemical potential $\mu$ of the set of $N$ molecules 
as a sum of a transverse part $\mu_t$ and a longitudinal part $\mu_z$:

\be
\mu=\mu_t+\mu_z
\ee

One  may  immediately  evaluate $\mu_t$ from  the  transverse  partition
function $Z_t$ because  that part of the problem  consists of independent
motions of individual molecules:

\be
-\beta \mu_t = \frac{1}{N} \ln Z_t = \ln \left( \sum_{\alpha} e^{-\beta 
E_{\alpha}} \right)
\ee

\noindent Here $\beta^{-1} = k_B T$ and the sum is over all 
eigenvalues $E_{\alpha}$ of the Schrodinger equation for 
transverse motion of a single
molecule. This equation may be evaluated analytically in the case when
the transverse potential is nearly harmonic, i.e. when $U$ includes just
a small quartic term, involving a parameter $\gamma$:

\be
U(\rho)=U_0 + \frac{k}{2} \rho^2 + \gamma \rho^4
\ee

The resulting transverse chemical potential is given by perturbation 
theory, assuming that the value of $\gamma$ is small:

\be
\mu_t = U_0 + \hbar \omega + \frac{2}{\beta} \ln (1-e^{-x}) + 2 \gamma 
\left[ \frac{\hbar}{m \omega} \coth \left(\frac{x}{2}\right) \right]^2
\ee

Here $x=\beta \hbar \omega$ with $\omega=\sqrt{k/m}$, and the regime 
of validity of this 
expression is that range of $T$ for which the rightmost term in $\mu_t$ 
is small compared to the middle term (i.e., $k_B T < k_B T_4 \equiv 
k^2/\gamma$). In 
the case of (10,10) nanotubes, the upper limit to the $T$ range 
for which this expression is valid is about 13,000 K (based on our 
fits to the Girifalco potential $U(\rho)$: $k=1.32$ eV/ \AA$^2$ and 
$\gamma=1.58$ \AA$^{-4}$). Thus, the expansion suffices in practice 
throughout the relevant range of $T$.

The present case corresponds to $x \approx$ 32 K/$T$, which is much 
less than 1 at room temperature. Therefore, it is appropriate 
to substitute the fully classical expression for $\mu_t$:

\be
-\beta \mu_t = \ln \left( \frac{1}{\lambda_T^2} \int \; d^2{\bf r} \; 
e^{-\beta U(\rho)}\right)
\ee

Here $\lambda_T = (2 \pi \beta \hbar^2/m)^{1/2}$ is the de Broglie 
thermal wavelength of the molecule, which is about 0.04 \AA$\;$ at room temperature. The resulting expression for $\mu_t$ in the case of a 
harmonic potential with a small quartic term coincides with the 
classical limit ($x \ll 1$) of the preceding quantum expressions 
for both $\mu_t$ and the transverse contribution to the specific 
heat (per molecule), $C_t/N$:

\be
\mu_t = U_0 + \frac{2}{\beta} \ln (x) + \frac{8\gamma}
{(\beta \omega^2 m)^2}
\ee

\be
\frac{C_t}{N k_B} = 2 - \frac{16 \gamma}{\beta (\omega^2 m)^2} = 
2 - 16 \; \frac{T}{T_4}
\ee  

This includes a constant contribution (expected for two harmonic 
motions in the classical regime of T), plus an anharmonic term 
proportional to $T$.

Figure 1 shows $\mu_t$ as a function of $T$, computed using the
 different aproximations described above. The full line is the quantum result (Eq. (6)) obtained assuming the nearly harmonic transverse 
potential of Eq. (5). It is interesting to note that the corresponding classical expression, Eq. (8), is extremely similar to the quantum one (they coincide in the figure), showing that quantum effects are really negligible. The dashed line shows the classical result Eq. (7) using the exact potential. For $T > 500$ K, the curves start to differ as the real shape of the potential deviates from the quartic aproximation.

The solution of the remaining 1D problem of interacting buckyballs 
requires a numerical calculation, in general. Quantum calculations 
have been carried out for some 1D fluids (helium and hydrogen 
\cite{boni,boro} in their ground states), but for classical systems at 
finite $T$ the determination of the equation of state is more 
straightforward. Specifically, the equation of a 1D classical system 
assumes an analytic form in the case when nearest neighbor interactions 
adequately represent the total potential energy \cite{1dnn}. 
We assume that to 
be the case here (and justify that in Section III) so that we may write 
down an expression relating the 1D density, $N/L=1/a$, in terms of the 
1D pressure $P$:

\be
\frac{1}{a}=\frac{\int_0^{\infty} dz \;e^{-\beta[V(z)+zP]}}
{\int_0^{\infty} dz \;z\;e^{-\beta[V(z)+zP]}}
\ee

\noindent where $V(z)$ is the C$_{60}$-C$_{60}$ interaction given by 
Eq. (2). Similar results were reported recently by Hodak and 
Girifalco \cite{thesis,press}.

Figure 2 shows the ``compression ratio'', i.e. the ratio of $P$ to 
the pressure of an ideal 1D gas  ($k_BT/a$); the interaction effects 
can be 
clearly observed. The pressure is evidently reduced 
by interactions over an extended density range until the point when 
the molecules start to come nearly into contact ($a \approx 12.5$ \AA
$\;$ at $T = 300$ K), above which density the pressure rises 
extremely rapidly. One observes that the C$_{60}$ gas behavior is 
nearly ideal at 
$T > 1000$ K while the effects of the interaction are quite 
significant below room temperature. The Boyle temperature is 
around 1400 K. The calculated behavior and high characteristic 
temperature are not surprising in view of the fact that the 
well depth of the pair potential is  about 3200 K. 

The 1D pressure can be integrated (according to the Gibbs-Duhem relation) to obtain the total chemical potential of the C$_{60}$ fluid at 
temperature $T$ and density $1/a$:

\be
\mu(T,a)=\mu_t(T) + \mu_{id}(T,a_0) + \int_{P_0}^P \;a(P',T)\; dP'
\ee

Here $\mu_{id}(T,a_0) = k_B T \ln(\lambda_T/a_0)$  is the chemical 
potential of a 1D ideal gas at a reference density $(1/a_0)$ that 
is low enough for the fluid to be ideal \cite{a_0}. The integration 
domain extends from the reference pressure $P_0=k_BT/a_0$ to the 
pressure corresponding to the specified spacing $a$. The result of 
this analysis appears in Figure 3 for a range of temperatures.
 
Figure 4 exhibits the effects of the interaction on the chemical 
potential, expressed in terms of the difference between the 
chemical potential of the (fully interacting) system and that 
of an ideal system which has the mutual interactions ``turned off'':

\be
\Delta \mu_{int}(T,a)=\mu(T,a)- [\mu_t(T) + \mu_{id}(T,a)]= 
\int_{P_0}^{P} \; \left( a(P',T)-\frac{k_B T}{P}\right) dP'
\ee

The rightmost integral may alternatively be expressed in terms of 
the 1D compressibility, $\kappa = - (\partial \ln a/\partial P)_T$ 
as follows:

\be
\Delta \mu_{int}(T,a) = \int_{\infty}^a \left(\frac{k_BT}{a'}-\frac{1}
{\kappa(a',T)}\right) da'
\ee

Because the attraction dominates the effect 
of the interaction over most of the density range shown in Figures 
2 to 4, the chemical potential shift due to the interaction is 
negative until one reaches high density.

\section{Corrections to the model}

We assess three corrections to the preceding section's results. 
One is the long range correction to the thermodynamic properties. 
This is the energy shift associated with the neglect of interactions 
other than nearest neighbors and could be written in terms of 
the probability per unit length $P_>(x)$ that a second, third 
or other (than first) nearest neighbor is present at distance $x$. 
Since we seek only an estimate of its effect, we take $P_>(x)$ 
to be a Heaviside step function for $|x| > 2 \sigma$, where $\sigma$ 
is the hard core diameter (7.1 \AA) of the C$_{60}$ molecule. 
Then, our approximate energy correction per molecule due to 
this neglect is given by

\be
\Delta E_>/N = \int_{2\sigma}^{\infty} \frac{dz}{a}\; V(z)
\ee

Note that a ``missing'' factor of (1/2) is cancelled by the 
omission of the region $z<0$. If the potential $V(z)$ were the 
familiar Lennard-Jones 6-12 interaction, the result of this 
integration would be $\Delta E_>/(N\epsilon)= - \sigma/(40 a)$ 
(where $\epsilon$ is the well depth), which represents a negligible 
correction  of the energy. 

The second correction we assess is the quantum correction. To do 
this, we introduce a phonon model as a means of characterizing 
the dynamics of the fluid. As was shown recently \cite{cross}, 
for a Lennard-Jones system in its ground state, the ratio 
 of the zero-point energy $E_{zp}$ to the pair potential well 
depth $\epsilon$ is

\be
\frac{E_{zp}}{\epsilon}= 2^{4/3}\;\frac{3}{\pi^2}\;\Lambda^*
\ee

Here $\Lambda^*=h/[\sigma (m \epsilon)^{1/2}]$ is the de Boer 
quantum parameter. In the case of a C$_{60}$ modeled by such 
an interaction \cite{LJ}, $\Lambda^* \approx 
0.003$, so this ratio is 0.002. This indicates that the quantum 
effects are small even in the ground state. They become even 
smaller (relatively speaking) with increasing $T$ and can be 
neglected at $T > E_{zp}/k_B$, which is of order 10 K for C$_{60}$.

Finally, we estimate the effect on the mutual interaction energy 
of the molecules' motion perpendicular to the axis. To do so, we 
evaluate the expectation value of the difference in potential 
energy due to changing the approximate separation of a pair of 
molecules ($z_{ij}$) used in Eq. (1) into their fully 3D separation 
$r_{ij} = (\rho_{ij}^2 + z_{ij}^2)^{1/2}$. To lowest order in 
the transverse separation, the difference in potential energy 
is $\Delta V(z) = V'(z) [\rho_{ij}^2/(2 z_{ij})]$, where the 
prime means a derivative. Substituting the mean value 
$\langle \rho_{ij}^2 \rangle=2 d^2$, the energy shift per molecule due to 
this substitution is obtained from the radial distribution function 
$g_{1D}(z)$:

\be
\frac{\Delta E_t}{N} = \int \; dz \; g_{1D}(z) \; \frac{V'(z)d^2}{z}
\ee

For this estimate, we substitute for $g_{1D}(z)$ a step function 
at $z=\sigma$. The result for a Lennard-Jones potential is

\be
\frac{\Delta E_t}{N \epsilon}= -\frac{24}{91} \; \frac {d^2}{a \sigma}
\ee

For a (10,10) tube in a high density case, with $a=\sigma \approx 9.5 \;$ \AA, for example, at room temperature ($d=0.5 \;$ \AA), we obtain 
$\Delta E_t/(N \epsilon)= 6 x 10^{-4}$, which is negligibly small 
(using $\epsilon \approx $ 3200 K). Although this is a relatively 
crude estimate, the order of magnitude is plausible in view of 
the small ratio of $d/a$. Evidently, the small value of $\Delta E_t$ 
implies that the 1D approximation is adequate for the present 
circumstances.

\section{Transitions due to interactions with neighboring tubes}

In this section we evaluate an interesting possibility- transitions 
of the C$_{60}$ fluid as a result of molecular interactions 
between buckyballs in neighboring tubes. In our analysis, we assume 
an infinite array of perfectly parallel and identical tubes and 
chemical equilibrium among the C$_{60}$ molecules. Under these 
circumstances, molecules in nearby tubes can undergo a cooperative 
transition, a condensation to a 3D liquid. A qualitatively similar 
problem (condensation of helium confined within interstitial 
channels of a carbon nanotube bundle) was studied previously by our 
group, using a Monte Carlo evaluation of a lattice gas model 
\cite{hecond}.

The present analysis is carried out first by perturbation theory; 
then, a comparison is made with an exactly solved model that 
captures the essential physics but employs a number of simplifications. 
Although neither model is completely satisfactory, we will see 
that the two sets of results are quite similar because the key 
predictions (parameters of the critical point) turn out to be 
only weakly dependent on the assumptions. 

The perturbation theory is based on the shift in free energy 
of the system due to interactions between molecules in neighboring 
tubes, $\Delta F_{inter}$ :

\be 
\Delta F_{inter} = \frac{N \nu}{2 a} \int_{-\infty}^{\infty} 
\,dz \; V[(b^2 +z^2)^{1/2}]
\ee

Here $b$ is the separation between tubes and $\nu$ is the number 
of nearest neighbor tubes. This expression assumes a simple 
form in the case when $b$ is sufficiently large that the interaction 
may be approximated by the asymptotic form $V(r) \approx - C_6/r^6$, 
in which case the integration yields 

\be
\Delta F_{inter} = - \nu N \; \frac{3 \pi}{16}\;\frac{C_6}{b^5}\;
\frac{1}{a} 
\ee

Including this term in the equation of state yields a relation for 
the shift of the 1D pressure proportional to $1/a^2$

\be
P=P_{1D} - \frac{\alpha}{a^2} \;\;\;\;; \;\;\alpha=\nu \,\frac{3 \pi}{16}\;
\frac{C_6}
{b^5}
\ee

Note that $\alpha=(3 \nu \pi/16) b V(b)$; thus it is proportional 
to the product of the strength of the transverse  interaction 
and the number of neighbors interacting with a molecule 
($\approx b/a$).

Figure 5 shows the effect on the isotherms of including the 
intertube interaction term. One observes that van der Waals loops 
develop, indicating the appearance of a phase transition. 
Figure 6 
shows how the critical temperature $T_c$  
varies with the parameter $\alpha$. For a typical lattice of (10,10) 
tubes with $b=17$ \AA$\;$ and $\nu=6$, $\alpha \approx 2180$ K \AA 
which yields $T_c \approx 530$ K. There is a singular rise of $T_c$ 
for small $\alpha$, followed by a region of weaker dependence. We 
may understand this behavior by examining an alternative model, the 
anisotropic Ising model applied to the lattice gas version of this 
problem. As shown by Fisher \cite{fisher}, for the case of such a 
model with 
nearest neighbor interactions $J_l$ along the 1D chains and 
transverse interactions $J_t$, the critical temperature is given 
by the expression

\be
k_B T_c =\frac{2 J_l}{\ln(1/c)-\ln[\ln(1/c)]}
\ee

\noindent where $c=J_t/J_l$. This is an asymptotic formula valid when 
$c < 0.1$ (weak transverse interaction).
Figure 6 shows the dependence derived from this expression when 
the values $J_l=800$ K and $J_t=18.2$ K are substituted in the Fisher 
formula. 
The result is seen to track the perturbation theory behavior well. 
This similarity is surprising, at first sight, because the 
perturbation theory is a mean field theory, neglecting density 
fluctuations in adjacent tubes. We rationalize this agreement by 
noting that 
the exact calculation yields critical exponents (3D Ising model) 
that are, in fact, different from those of the mean field perturbation 
theory. On the other hand, in 3D, one does not expect a large error 
in the mean field theory value of the critical temperature 
\cite{3d}.

Finally, we note that Carraro has described the possibility of a 
crystallization transition for such a problem \cite{carraro}. 
To estimate the 
onset temperature for this freezing, we simply evaluate the energy 
difference $\Delta E_f$ between the ground state crystal and the 
energy of the alternative fluid state. This is determined by 
comparing the energy computed above ($\Delta F_{inter}$ ) for the 
intertube interaction energy. The difference is just the difference 
between the integral and the sum that arises when the buckyballs 
are located in lattice sites, separated by $a$:

\be
\Delta E_f = -\frac{\nu C_6 }{2 a^6} \left[ \sum_n \frac{1}
{(\delta^2 + n^2)^{3}} - \int_{-\infty}^{\infty} \frac{dz}
{(\delta^2 +z^2)^{3}}\right] 
\ee

\noindent where $\delta = b/a$.
The difference would be identically zero if the Euler-MacLaurin 
theorem were analitically used to evaluate it. The integral can be analitically 
evaluated yielding $3 \pi/(8 \delta^5)$. Hence, we compute the sum 
numerically and find the difference. It is negligible small when 
$b/a > 2$ and increases as $b/a$ goes to zero. This is shown in Figure 7, 
expressed as a freezing temperature

\be
T_f = \Delta E_f /k_B
\ee

In the case of C$_{60}$ inside (10,10) tubes, $b/a = 1.7$ and $T_f \approx 
0.5 K$. Such small values of $T_f$ imply that such a transition will be 
difficult to observe in the laboratory.

\begin{acknowledgments}

We are grateful to Victor Bakaev, Mary J. Bojan, Karl Johnson, Carlo Carraro, Louis Girifalco, Miro Hodak and David Luzzi for stimulating and helpful comments and to NSF and the Petroleum Research Fund of the American Chemical Society for their support.

\end{acknowledgments}

\begin{table}
\begin{tabular}{|c|c|c|c|} \hline \hline
      System      &$\;\;|\Phi(r_0)|$ (eV)$\;\;$&$\;\;r_0$ (\AA)$\;\;$&$\;\;r_1$ (\AA)$\;\;$\\ \hline
\,\,C$_{60}$-C$_{60}$\,\,&    0.278             &      10.05    &     7.10      \\ \hline
\,\,C$_{60}$-(10,10) tube\,\,&    0.537      &   13.28        &     10.12    \\ \hline\hline
\end{tabular}
\caption{Parameters of the universal function Eq. (2) for the carbon structures investigated.}
\end{table}

\newpage

\begin{figure*}
\includegraphics[height=4in]{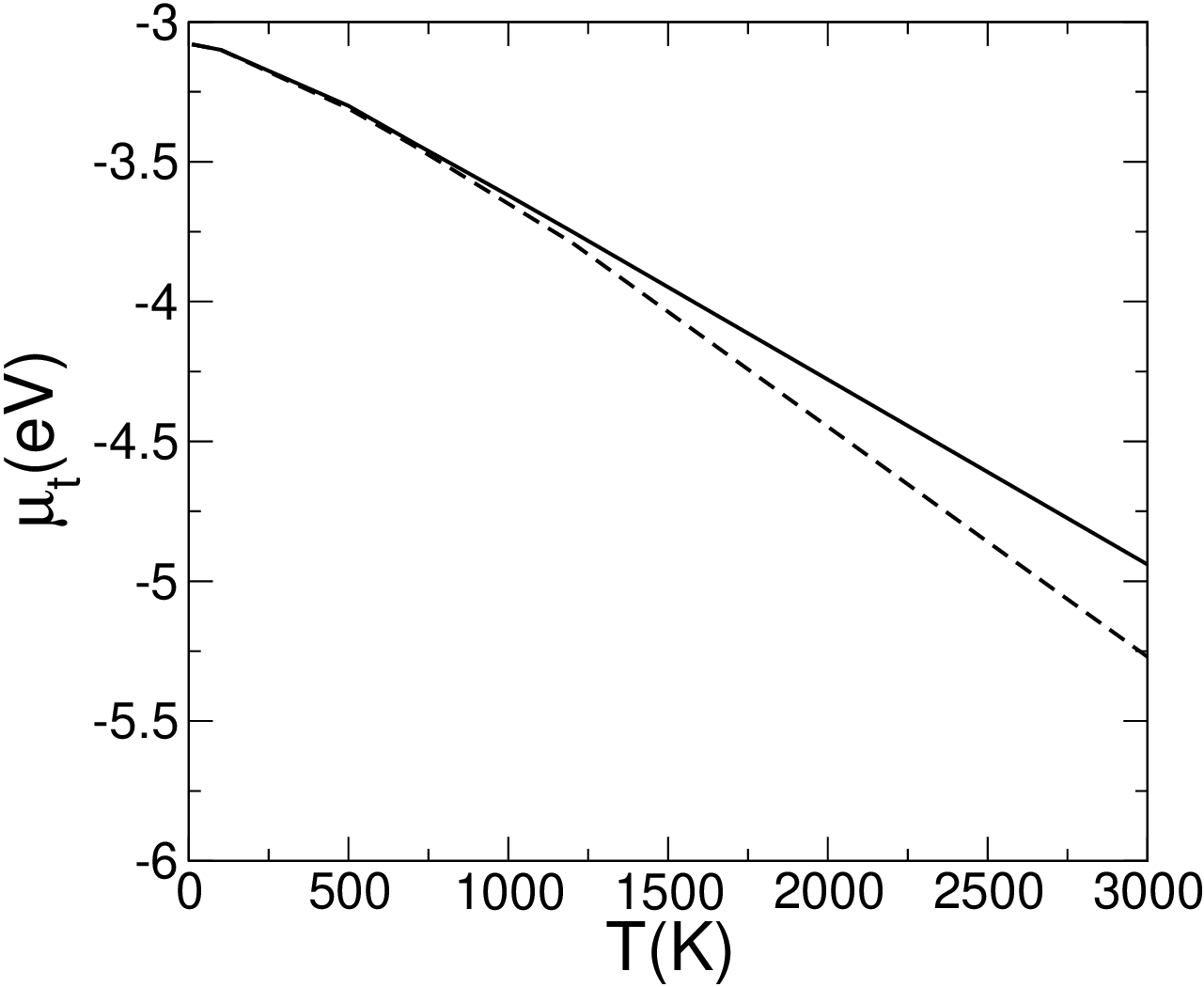}
\caption{The transverse chemical potential $\mu_{t}$ as function of temperature. The full line corresponds to the quartic aproximation of the potential (Eq.5); quantum and classical results (Eqs. (6) and (8)) 
are so similar that they cannot be distinguished over the whole range of $T$. The dashed line shows the classical result using the full potential, Eq. (7).}
\end{figure*}

\begin{figure*}
\includegraphics[height=5in]{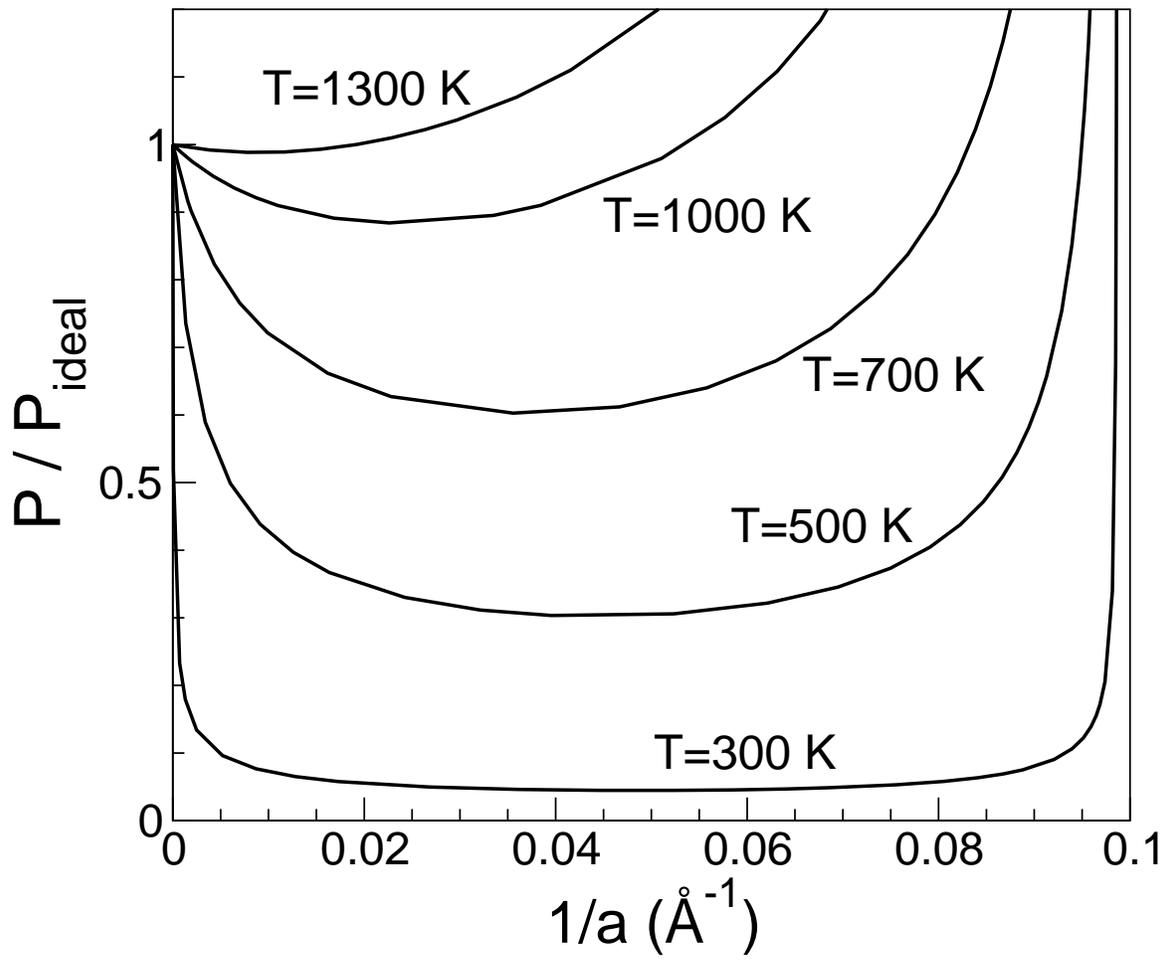}
\caption{Compression ratio at various temperatures of the 1D 
pressure $P$ to that of an ideal 1D gas, $P_{ideal}= k_BT/a$, as a 
function of the density $1/a$.}
\end{figure*}

\begin{figure*}
\includegraphics[height=5in]{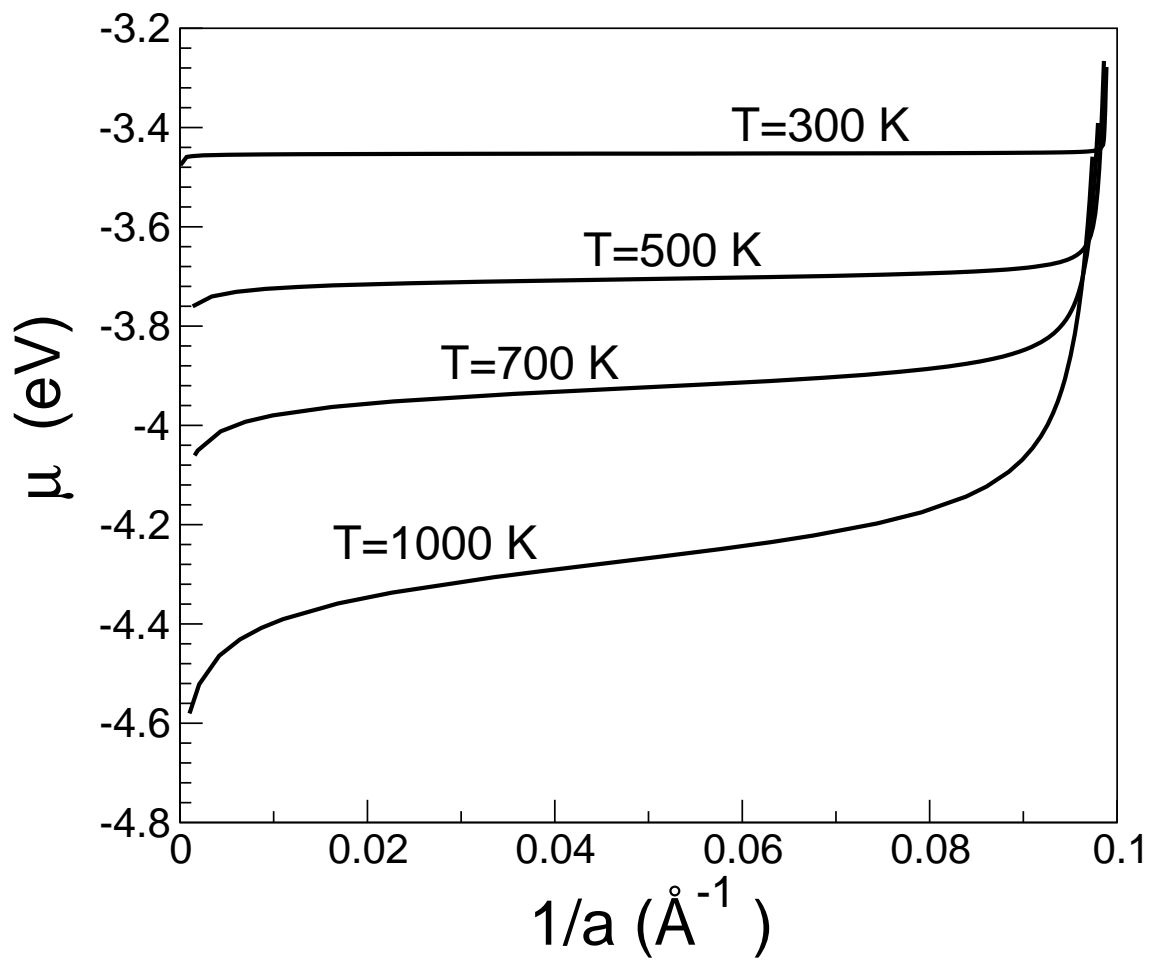}
\caption{Total chemical potential $\mu(T,a)=\mu_t(T)+\mu_z(T,a)$ 
of the C$_{60}$ fluid as a function of molecular density $1/a$ and 
temperature $T$.}
\end{figure*}

\begin{figure*}
\includegraphics[height=5in]{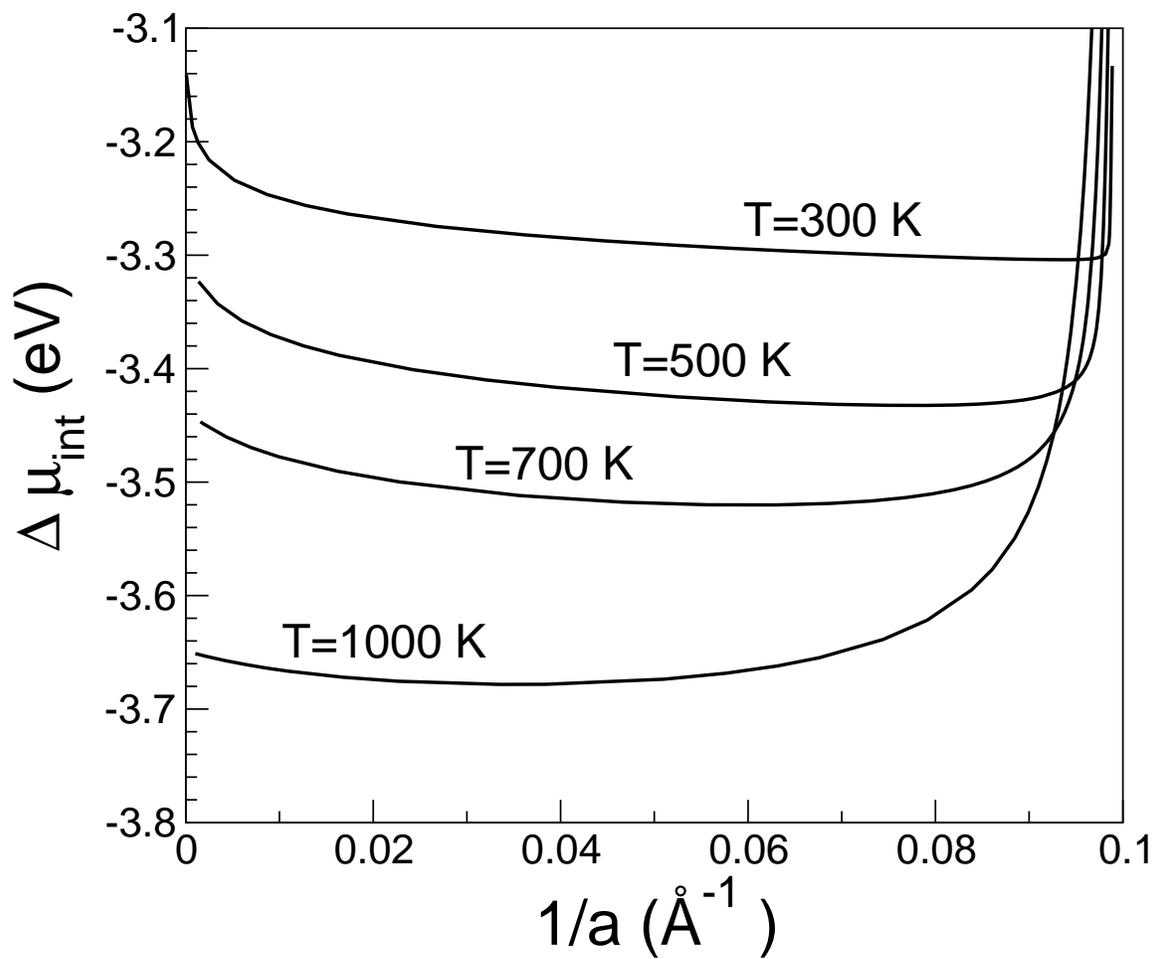}
\caption{Chemical potential shift due to the intermolecular 
interactions $\Delta \mu_{int}(T,a)$  as a function of $1/a$.}
\end{figure*}
 
\begin{figure*}
\includegraphics[height=5in]{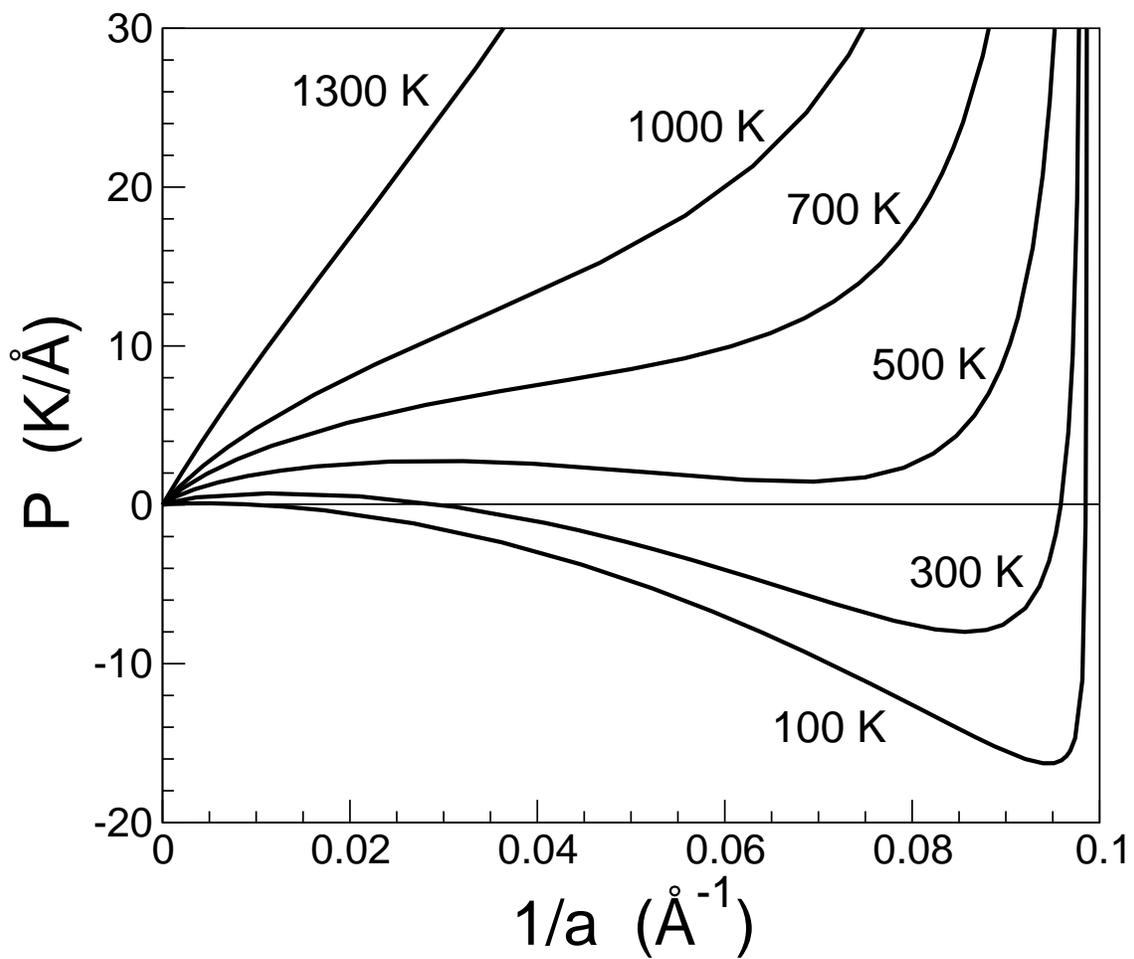}
\caption{Effects of the interaction between C$_{60}$ molecules in 
nearest neighbor tubes ($\alpha = 2180 $ K \AA). The isotherms 
develop van der Waals loops as a function of the temperature, with a
critical temperature around 530 K.}
\end{figure*}

\begin{figure*}
\includegraphics[height=5in]{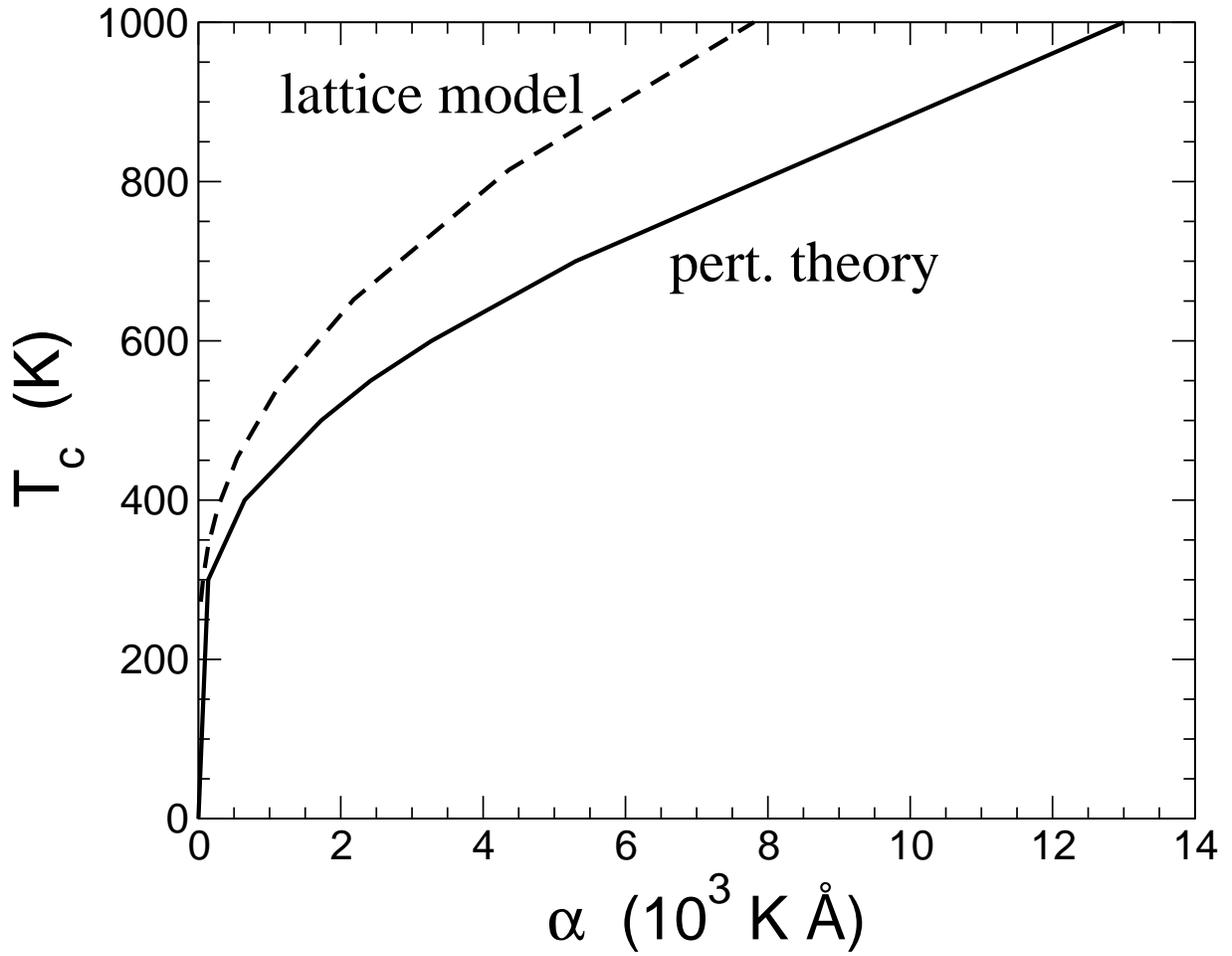}
\caption{Dependence of the critical temperature $T_c$ on the strength
 of the transverse interaction and geometry of the tube lattice, given by the parameter $\alpha$. The full curve corresponds to the 
perturbation theory results and the dashed one indicates the anisotropic lattice model predictions (see text).}
\end{figure*}

\begin{figure*}
\includegraphics[height=5in]{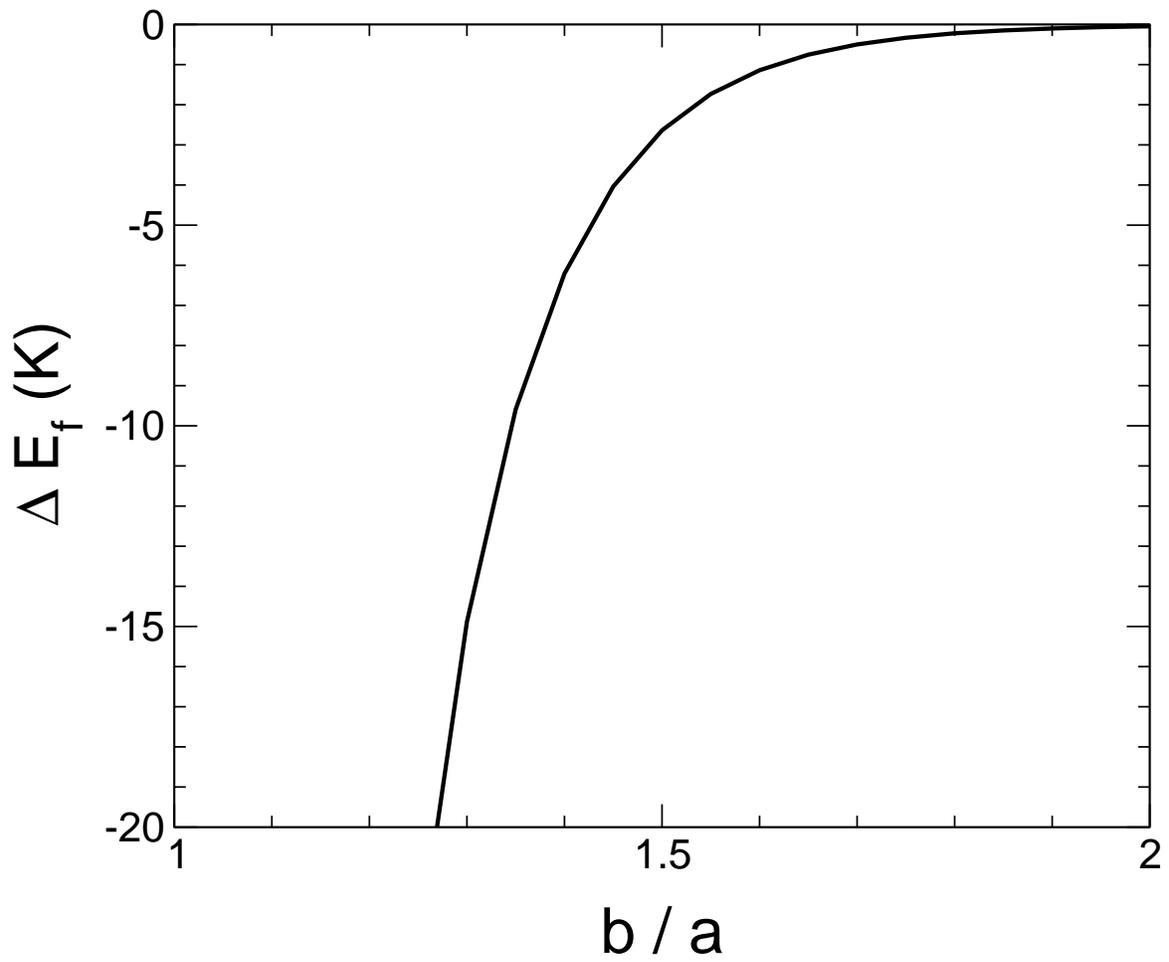}
\caption{Energy difference $\Delta E_f$ between the ground state 
crystal and the fluid state as a function of the dimensionless 
separation between tubes $\delta=b/a$.}
\end{figure*}


\begin{references}
\bibitem{a_0} The choice of this reference is arbitrary because the 
$a_0$ term cancels out after the integral is performed.
\bibitem{LJ} If a Lennard Jones type of potential was used to characterize 
the $C_{60}-C_{60}$ interaction, the parameters would be: 
$\epsilon \approx$ 3200 K and $\sigma \approx$ 9.5 \AA.
\bibitem{3d} For the 3D Ising model, the mean field theory's error 
in $T_c$ is $\approx$ 33\%.
\end{references}
\end{document}